# Implementation of Portion Approach in Distributed Firewall Application for Network Security Framework

Harleen Kaur[1], Omid Mahdi Ebadati E.[2] and M. Afshar Alam[3]

[1] Dept. of Computer Science, Hamdard University
New Delhi, 110062, India

[2] Dept. of Computer Science, Hamdard University
New Delhi, 110062, India

[3] Dept. of Computer Science, Hamdard University
New Delhi, 110062, India

## Abstract

The stimulate of this research seeks collaboration of firewalls which, could reach to the capability of distributed points of security policy; the front-end entity may much interact by the invaders so the separation between this entity and back-end entity to make the secure domain protection is necessary; collaborative security entity has the various task in the organization and there is a certain security policy to apply in; the entities like DPFF have to be protected from outsiders.

Firewalls are utilized typically to be the main layer of security in the network framework. The research is presented the particular segment of the proposed framework that DPFF based on the developed iptable firewall to be the layers of defense, which is protected front and backend of the framework with a dynamic security and policy update to control the framework's safeguard through proposed portion approach algorithm that utilize to reduce the traffic and efficiency in detection and policy update mechanism. The policy update mechanism for DPFF is given the way of its employment.

The complete framework signifies a distributed firewall, where the administrator configures the policy rules set, which could be separately or else from administration nodes' side.

## 1. Introduction

A firewall is a hardware or software for defending the privacy, reliability, and accessibility of income and outcome packet over the network. Firewall expertise has enhanced significantly since it was introduced in the early 1990s [4]. The premature firewall worked with simple packet-filtering firewalls, but it has been developed with more capability to rely on multiple layers of network activity. As the World Wide Web has developed into the progressive and complicated network of today, the Internet and network security [5], has become more problematical, with various attacks and break-ins. Nowadays firewall technology is a typical part of organizations to provide security of networks [2].

In actual fact, today's firewalls present a security fence between any networks, where the flow of traffic requirements to be controlled and observed [21]. To reduce network security risks, appropriate network access policies must be defined as first defend layer of defense in organizations' strategy. Firewalls implement such policies. Firewall deployed polices traffic flows between internal and external networks. The firewall application is to be a segment of typical security strategy, which cannot be lonely a most complete secure area [27]. These, such technologies essential are complemented with other security technologies that may provide a complete solution.

Rules set policy let the firewall either permits or denies access to the network. Thus, a firewall may be placed to authorize all confident traffic and to deny all other requests, and in addition it may also set to deny all messages of a particular kind except of specified network addresses or IP domiciles. Firewalls are intended to be a safeguard adjacent to effective endeavors of an assortment of security breaches. Firewall roll is to focus on security management [20], at a certain point; by this means abridge the accomplishment of security policy, the pathway of information and sometimes auditing. A firewall can also collect attack substantiation [10], to permit an organization to follow officially authorized action.

There are limitations for the firewalls which unable them to defend hostile to malicious [25], insiders or unknown attacks or threats. In addition, firewalls are not mainly deliberate to defend against virus attack but some models propose to work as virus detector or protector to perform in arbitrary data packets passing through a firewall.

## 2. Background

Firewall administrator typically, is located within the network administration to organize the services and give the individual effort to be comprehensive of policies and rules establishment in an organization. Based on a firewall definition by W. R. Cheswick, et al. [7], which design a firewall is for controlling all inside and outside traffic and just through it and traffic based on the authorization local security policy will only be allowed to pass and firewall itself should be protected and unaffected to penetration, and have a task with the collection of elements situated between two networks to conduct the mentioned properties.

A research by E. D. Zwicky, et al. [29], has a broad description of firewalls fundamental, that they can be a secure checkpoint for inside and outside of a connection. The hosts available on a network could have any number of ordinary servers for various functionalities, so the lowest cost option is monitoring the router based on sophisticated fashion. A direction for multiple part of network with firewall control, which has the security restriction and implementing policy for multi-part, is given by J. D. Guttman [12]. In extension for different environment and have the distributed model which networks host may be employed in different portion and control by organization security policy is presented by S. M. Bellovin [3], S. Ioannidis, et al. [17]. However, M. Bellovin [3] has not given any report for implementation of this method. His approach proposed a few features of IP firewall functionality at the endpoints of the communications. One of the merits on this approach is employment of decision making at the certain location and receiving more information than a router through the firewall.

A research by S. Ioannidis, et al. [17], focus on trusted management in the system and by modification of BSD kernel to support the implementation of firewall control at TCP connection, which was disregarding the router based firewall idea on decision making and an approach through the message header, but their goal was not a particular work on network environment for fully security structure and deployment.

An enterprise or organizational security policy is undertaken and develop regards to threats [26], which the desirability and outcome and costs issues considered by M. Coetzee, and J. H. Eloff [8] and D. Moore, et al. [22]. Security's analysis, which is a set of decision for permitting or denying different elements on a network based on the scope of organization is made. The setup of a firewall software or hardware is requiring a policy set on each node and implementation of that. As defining the policy for network control is important logging of this activity is much important.

An automate development of policy by M. Hamdi, and N. Boudriga, [14], is developed different techniques in computer security by using attack graphs. The work is attempted for growing very large attack's graph through pruning and management. The network policy tool was expanded by J. D. Guttman [13], which carried out on generation of route set as well as large filtering. This method utilized logical statement for required behavior on each node between the areas and truly implements the entire rule sets.

To improve the security policy and protection of the Internet browser is presented by L. Huang et al. [15], is research on various types of browsers attack such as the network abilities' interaction with client, threats scopes, network attacker, XSS phishing, malware, weak authentication and user tracking to control a web attacking by using a validate domain, certificates like SSL, HTTPS to prevent network attacks and also implementation difficult practice C. Jackson, and A. Barth [18] and C. Jackson, and A. Barth [19], on resources. However this paper presented in our main research proposed model as the front end layer of the security in collaboration with Distributed Network Intrusion Detection System [30], [31] which located behind the Distributed firewalls to be two main components of the main research.

## 3. Analysis of Firewall Activity

In firewall activity, there are several queries about its activity and traffic control. Some of these enquiries are as follows:

- Particular source and target and their services which may be accessible by source should be investigated.
- Comparison between two distributed firewall for their configuration to have the corresponding enforce same policy.
- Investigation about the active firewall and if one of them be active.
- The influence of a node compromises to interface the firewall.
- Investigation about the conducted policy configuration to meet the organization requirement.
- All the open ports that may not to be open to inbound or outbound of the available nodes should be disabled.
- The ports according to the organization policy, which require communicating, should be defined in firewall policy.
- Correctness of the rules updated on the bases of organization policy should be examined.

With the appropriate examine procedures the firewall policy utilization base on its specification is interested to use and coincidence and conflict in policy should be indicated to the administrator.

Our aim is an analysis that is based on applying the formal requirement to define the firewall policies and their relations, which assists to identify probable errors and firewall policies weaknesses aforementioned to their development in the framework. The store policy satisfies the conditions can be as following forms:

*If source IP and port and*
*If target IP and port then*
*Do exploit*

In the firewall filtering situation various conditions to satisfy the policy typically refers to equivalency between present packets of the current connection and the store policy which worked according to the applied algorithm. In our propose framework, every rule for the firewall is consists of six characteristics, which are in the following layout:

< "Connection Protocol" "Source IP" "Source Port" "Target IP" "Target Port" "Action">

First part of packet is the protocol like UDP/TCP/ICMP. The next two parts are referred to IP domicile and port of source and also the next two parts to the source are the IP and port target of packet. Lastly, the final part is the action which decision of the firewall in upon it.

Here we define the policy for the firewall activity that can have action according to the correlation of packets.

The decision of the firewall to bypass the packet or drop is based on the comparison of the network fields and depends on the defined policies. The relation in the policy is defined by $P_m$, which contained all of current defined policy, which could be updated through $P_n$, which is the new rule that should compare with current policy by the using the analyzer. The result of comparison could be completely matched, almost match, incompletely match, completely disjoint and interrelated. The following example can show the correlated between $P_m$ and $P_n$ , and as it is shown may be some fields of $P_m$ are equal or subset to the equivalent of $P_n$, and also shows that the $P_m$ policies are superset in comparing to $P_n$.

***Example:***
*If a, b, c $\in$ {Connection Protocol, Source IP, Source Port, Target IP, Target Port} where a $\neq$ b $\neq$ c*
*And also $P_m[a] \subset P_n[a]$ or $P_n[a] \subset P_m[a]$ or $P_m[a] = P_n[a]$ then*
*$P_m[b] \subset P_n[b]$ and $P_n[c] \subset P_m[c]$*
*That Implies:*
*The $P_m$ and $P_n$ have interrelated policy.*

*<TCP Protocol>, <Source IP 124.125.1.15>, <Any Port>,*
*<Any Target in Framework \*.\*.\*.\*>*
*<8080>*
*<Not Permit Drop>*

The firewall activity and update filtering mechanism could be done based on the following flowchart (Fig. 1).

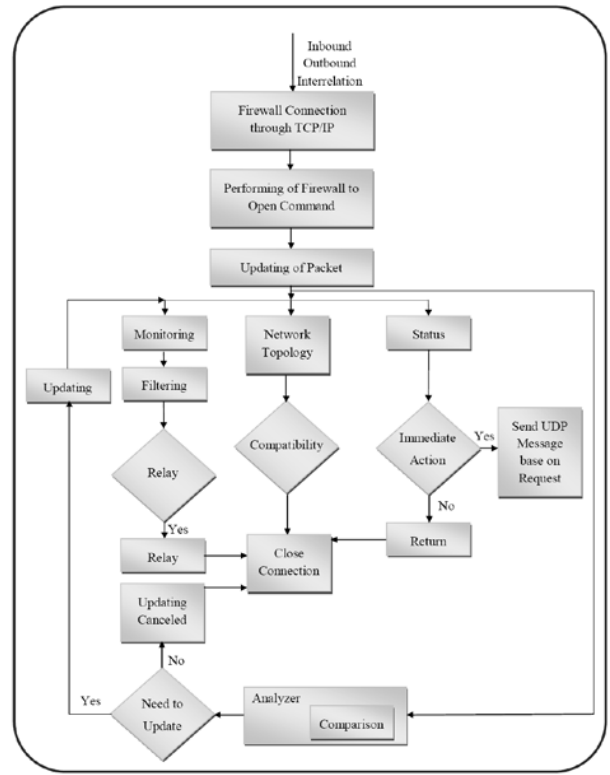

Fig. 1 Proposed Firewall Activity and Update Filtering Mechanism Flowchart

As it is shown in firewall connectivity of TCP/IP after receiving the packets (inbound and outbound) firewall performs to open command to receive and initial its investigation. In next firewall is starting to validate the packet based on its policies and somehow updating. In the parallel activity, firewall check for monitoring the network topology and its status. For the first part after monitoring the packet it will check for filtering, relay or update, in the second, the network topology it checks for capability of this topology, and in last, the status for immediate action and response is checked.

## 3.1 Firewalls Limitation

To briefly explain the limitation of the firewall as a front and backend first layer of defense in proposed security should mention that:

- Generally, a threat by using a mobile host communication can pass the firewall protection.
- Internal treats cannot control by the firewall.

Intrusion activity is not under control by the firewall.

## 3.2 Access Control

The main requirement to define a secure access control is to have trusted systems:

- Environment Access: The method to support and guide the framework from intruders is to instigate trusted system technology like data access control that authenticated the users' access and controls them.
- Multi-Security Point: By defining several categories and points of data and users can be defined to access and the obligation in subject to at the high level of security the defined data level and user access is non comparative to the lower level, unless the valid user is authenticated. In this technique two parameters should be employed, first is accessing to the level of security, but only can read the available object, and second the object can modify and access to write in case the access permitted.
- Monitor Orientation: on the bases of security parameters that define with framework policy the elements are control in the operating system and hardware. Security of the operating system kernel is an important matter to safeguard the framework and the next is monitoring the access privileges for each object and characteristics of protection. The security enforced the policy to conduct two previous property for all entities to reach to the following:
  - Security's policy should be enforced for each access to the media.
  - The parser and analyzer due to the orientation monitoring will be safe from the modification or unconstitutional approach.
  - The testimony of the framework may observe by distributed administrators.

## 3.3 Demilitarized Zone in the View of Firewall

To the view of firewall position that applies in various cases to the requirement of allowing external access to some part of the proposed framework such as the mail and web activity the concept of a secure zone, which define as DMZ, is defined. The purpose of the DMZ is to supply the network segment which is externally accessible, and also internal services are provided to the public view. By this purpose, the other segments of the framework are separated from the main DMZ and have their own activity, so in this method unauthorized users are not permitted from accessing public resources.

In circumstances of proposed framework servers are located on the same network segment behind the front firewall and before the backend firewall, and all inbound traffic is restricted as prior define. The main server as well as other servers are placed on the detach DMZ and connected to the firewall. The firewall policy for the security is set to permit all inbound connections through the port number 25 for the mail server located between two firewalls. In a worse case that an invader misused to reach to the administrator access, they may install a program for the full access through the back door software to the server. Nevertheless, since the servers as well as the mail server is not on the similar network segment as the other resources are and protected by the DMZ the hacker activity cannot be manipulated. As it has been explained in charter 6, the lure and interactive insecure segment of the network are always active specifically points the invaders to that parts of network.

Operating a public main server over the Internet is hazardous [7]. However, in the pervious example the DMZ reduced the risk and does not let the server compromised with the aim of DPFF. It is very important at the time of conducting the DMZ the public view of servers does not have any connection like a back door to the internal network, which in the proposed model is again become secure by the backend firewall, otherwise invaders could bypass the DMZ and inter to the client and protected internal corporate network, and the individual associate between the subnet and the conducted DPFF.

To provide beyond the security level multiple DMZ is offered in the proposed model. As per for the security reasons the following parameters are required to consider:
- The DMZ should not connect directly to the internal network (to the other network segments), and it should be isolated.
- As it has mention above the DMZ should not include the internal framework information that may be misused by the invaders.

# 4. Feature Selection Approaches

## 4.1 Characteristics of the First Layer Defense Security of the Proposed Framework

The intended approach for the firewall application is sorted as (Fig. 2):

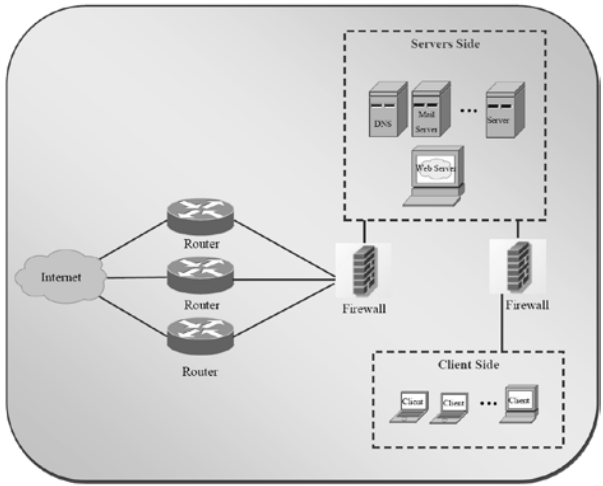

Fig. 2 Proposed Model Strategy for the First Layer of Defense

- All the network traffic (inbound and outbound) is essential passed through the firewall, which helps to obstruct all approach to the network except through the firewall (ISO/IEC 17799:2005, 2005).
- Only legitimate traffic, which is defined by the restricted security policy maybe permitted to bypass the firewall.
- Maintain the firewall protection to saturation with the utilization of a conviction system with the secure operating system.
- Network packet filtering on the bases of IP domiciles, ports number and specified protocol.
- Determination to permit or deny the network flows from the direction of the interest.

## 4.2 Demonstration of Policy Mechanism

The firewall investigates and checks every packet protocol and IP domicile information and after that it filters the outbound and inbound packets based on the set of security configuration policies. Accordingly, permission or dropping the packets are as the example of the following typically rules as well as particular policies, which may be updated through the analyzer and by administrator.

- It drops all network packets, which may be subject to updating policy or administrator commands.
- Limitless access to the web-server, which based mostly on port number 80.
- Limitless access to port number 25 to access mail-server through SMTP protocol.

A firewall policy commands how the network traffic bypasses through firewall. The applicable policy also illustrates the firewall updating and restriction. To create the firewall policy various risk analyses should be carried out on the bases of a requirement for the execution of the organization. By an overview of this analysis list of applications and how it shall be secure is processed. With the knowledge based of the vulnerabilities, which is associated with each application, the particular method will be used for securing this application. The risk analysis should estimate according to the infrastructure but in the propose model the evaluation has been done the various elements like, vulnerabilities, threat and invader activity and with the impact of sensitive data available on the servers. By this goal, the evaluation of these elements aforementioned to determine the firewall policy is analyzed and the structure of handling applications traffic is processed. The particular of those applications, which could pass through the firewall and the activities, can take place by involving the following steps to create the firewall policy.

- Recognition of vulnerabilities related with the essential application.
- Analysis of the approach model for securing the applications' investigation of protection method to create application traffic.
- Establish of firewall policy is to support the application traffic.
- Establish of firewall rules based on the IP domicile, ports and protocol.
- Identification of the various conducted system vulnerabilities to update the policies.
- Collaboration with other joint defense layers to completion of the policy.

By referring to the nature of academic environment, the campus firewall cannot be setup to characteristics industry general standard. Due to an individual requirement of the security and depends on kind of data to be protected and confined the proposed model typically follow the maximize the network security environment to utilize the minimum risk. On this bases the requirement of the blocking the traffic is conducted on the followings:

- All inbound traffic included ICMP traffic.
- All inbound traffic from a source, which is nota valid source system with a targeting of the firewall address.
- All outbound as well as inbound from a source system which may be in the range of private networks that may include in classes A, B and C of the network domicile.\
- All inbound traffic which maybe generated at the behind hand of the firewall that initiated in inside the proposed framework.
- All inbound traffic that may contain SNMP network traffic protocol from an invalid source.
- All inbound and outbound traffic which addressed to publicize the inside framework addresses.

- All inbound and outbound traffic that included the target or even source IP domicile of the local host.
- All inbound traffic, which may include the routing information of resource IP domicile.
- All inbound and outbound traffic that may include to target or generated from a source of non-IP domicile (0.0.0.0), which is the non-specified IP.

As the firewall logs are showing in Table III to protect the proposed framework from the firewall view following notes also needs to be consider:

- The proposed firewall model is to be implemented on a system which should not be occupied by the redundant applications and itself should cynical pastille to assail, by the means of all the security patches, which may be required for the security purpose in Linux is utilized and always is under take by the administrative view.
- The distributed packet filtering firewalls need particularly backup, which not be supposed to on any inside or outside of the proposed framework client or server which itself can be a perspective of the security hole to the proposed framework.
- The synchronization mechanism between the DPFF should be appropriately done and the consideration of their log activity on the time based is required.

The distributed firewalls are supposed to block all inbound traffic except that traffic, which is clearly required to inbound with the server connections, which are secured by the demilitarized zone. The Table I is shown the generalized policy to manage the inbound and outbound network traffic.

- In addition to above notifications the validation of the procedure and examinations to the stored logs and latest policy activity is required to minimize the unknown incident.
- In SMTP for the mail server activity, the following table (Table II) content should be considered [22], and block to avoid the internal explosion, which typically should define on remote antivirus as well.

Table 1: Sample of Policy Rules Based on The Protocol

TCP, INPUT, 10.0.0.3, 139, 121.10.5.3, 49621, ACCEPT

DNS, INPUT, 10.0.0.1, 53, 127.14.41.11, 61190, ACCEPT

TCP, OUTPUT, 209.85.231.100, 49625, 10.0.0.4, 80, DENY

UDP, OUTPUT, 255.255.255.255, 54507, 10.0.1.12, 2223, DENY

UDP, INPUT, 10.2.0.4, 51005, 10.0.0.1, 53, ACCEPT

TCP, OUTPUT, 229.96.11.79, ANY, 10.14.7.3, ANY, DENY

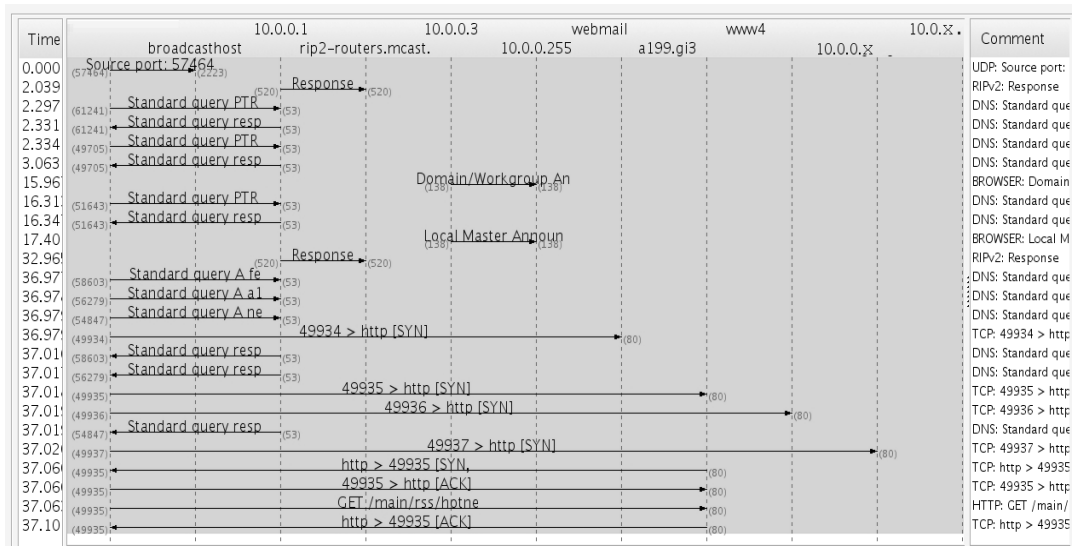

Fig. 3 Sample of FlowGraph of SYN/FIN/ACK During Experiment

Table 2: File Extension for Smtp Exchange

| File Extension for SMTP Exchange | | | | | | | | | |
|---|---|---|---|---|---|---|---|---|---|
| . acm | . ade | . adp | . ax | . bas | . bat | . bpl | .bss | .ca | .cc |
| .cfa | .ch | .chm | .cmd | .com | .cpl | .crt | .dll | .do | .dpl |
| .eml | .ep | .exe | .fc | .fffa | .fnfa | .ga | .gcda | .gf | .gfaa |
| .gfs | .gpa | .hlp | .hta | .inf | .ins | .isp | .js | .jsa | .lla |
| .lnk | .mdb | .mde | .msc | .msi | .msp | .mst | .ocx | .pcd | .pif |
| .pl | .pot | .rdata | .reg | .reloc | .rf | .rva | .scda | .scr | .scf |
| .scx | .sfaa | .sfp | .shb | .shs | .sys | .url | .vb | .vbe | .vbs |
| .wf | .wsc | .wsf | .wsh | .xl | | | | | |

## 4.3 Implementation of DPFF Policy

It depends on the proposed framework architecture the following policy is implemented as the minimum requirement:

- The source address of the packet which initiated from layer three should be generated from an IP domicile.
- The target address of the packet which initiated from layer three should be generated from an IP domicile.
- Based on the forth layer communication, inbound and outbound packets should contain correlated ports and protocols for the source as well as the target. Fig. 4 and Fig. 5 are shown various protocols and their related percentage that contained in the experiment.
- The inbound and outbound traffic from the source/ to a target regularly could be Ethernet in layer two and also IP at layer three, which shows the type of traffic.
- To collect all the activity logs naturally then the packets are dropped are not logged by the firewall

but in the proposed model all the activities are recorded as well as denied and dropped action.

The firewall policy can be accumulating after implementation the applications traffic and based on the firewall the network traffic controlled. To avoid the unintentionally firewall bad activity or whole policies are simple as much as is possible that also help in unauthorized firewall traverse traffic.

## 4.4 Proposed Policy Algorithm

In consideration of actively work on the traffic and have the best performance of the firewall and also optimize the traffic at least an amount by optimizing the algorithm [11, 24], and the requirement to classify the various relations and connections among the firewall rules and policy [28], and to match all the traffics by investigation in between the rules and policies based on the following definition and activity the respective process is generated.

Each portion of the traffic is defined as the split of the whole amount of traffic, so for each portion every header packet corresponded precisely to the similar collection of the rules and also neither any other associated part of the packet nor even the packets can have this similarity for this particular rules' collection. This is by mean of every packet is only fitted into a similar portion, and it is impossible to tell apart from the other inspection of the whole firewall policy having the similar header.

To define these portions, it should mention that the included rules to apply for the particular space is defined as $R_{in}$ and similarly the exclude rules are $R_{out}$, also $AddSp$ is in place of Boolean expression of the address space and finally *action* of the firewall for that particular space based on the initiate rule to the end of the list indicate by action.

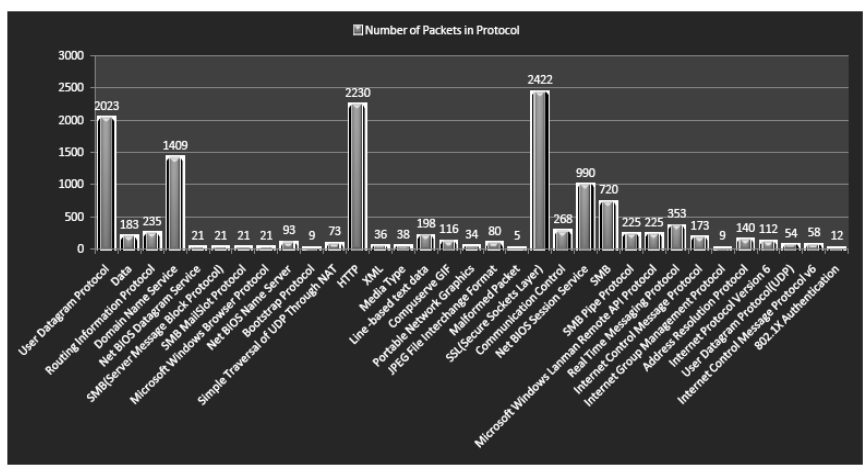

Fig. 4 Number of Various Packets in Protocol

To clear the viewpoints of this algorithm, which assemble a record of portions that is based on the policy interface and communication is each time after acknowledgment of the portion it initiated at the portion as the input and interconnects through the $R_{in}$ to the effectual rules of the policy ($R_{eff}$). In this situation, there is a condition that, in case the included rule has the unfilled record then the *action* to this portion is considered as default and in the reverse case it will be the base on the priority of the rule and based on their arrangements. In this algorithm (Fig. 6), the *For* loops are responsible to oblige and influence of the whole portions and calling them to be included in the loop. The include and exclude portions, *IncludePor* and *ExcludePor* are represented for the present procedure of the portions to include or exclude between the rule and the portion, and also somehow break portion into the contain or un-contain regions, which mean by abundant the portion as the unbroken space for the excludes the portion in case the portion and the rule are not interconnected, but in the reverse case for the include the rule will be covering whole of the portion.

Here, there is the necessity of the provision situation for the portion to avoid of the unfilled portion construction. Finally, based on this expansion the algorithm can have the result of doubling of the portion subsequent to dealing out of the $n$ rules of the policy and consequential has the record of $2^n$ portions based on this escalation. This algorithm is based to the continued algorithm and wisely chose the packets and covers the whole space of the packets in the proper way. Typically, by utilizing this algorithm the address space divided into portions according to the rule, and each portion takes the particular address space and based on the priority each portion is considered. So, the rules of policy from every portion are pulling out, and the production and outcome will log to consider for the firewall, and the complexity is resulted as O($2^n$).

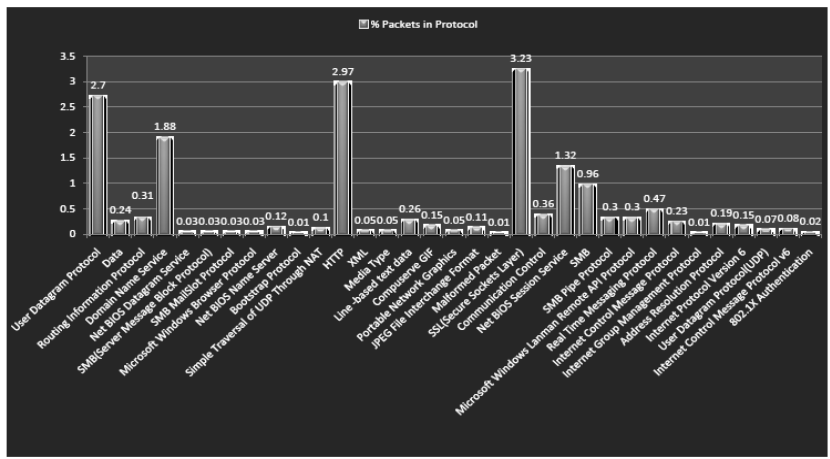

Fig. 5  Percentage of Various Protocols in Firewall Experiment

```
Input PLIST ← P
AddPor (Initial Domain, P, P, DefAction)

For all Rules r=1 to n do
For k=PLIST-1 to 1 do
G=PLIST[k]
IncludePor ← G.AddSp ∩ AddSp(R_r)
ExcludePor ← G.AddSp ∩ ¬AddSp(R_r)
If IncludePor != Por.AddSp then
Portion Is Not Contained in Rule's AddSp
If IncludePor != Φ then
AddPor(IncludePor, G.R_in U {R_r}, G.R_out, G.R_eff U {R_r})
Else
No Intersection of Rule and Portion
AddPor(ExcludePro, G.R_in, G.R_out U {R_r}, G.R_eff U {R_r})
Else
Portion is Inside the Rule's AddSp
AddPor(IncludePor, G.R_in U {R_r}, G.R_out, G.R_eff)
PLIST.Delete (Portion k)
Return PLIST
```

Fig. 6 Portion-wised Rules for Policy Algorithm

## 4.5 Policy Configuration

Currently there is a lack in firewall configuration to be as a semantic and a typical language define. Based on the various unique configuration in high level configuration language, which proposed to be a semantics in firewall languages for access policy that mostly working on access control policy in network management these mis-configurations are clearer [10]. In the proposed model, the firewall configuration to be interpreting in semantics is working and configured by XML syntax, which is commanded to the high-level access control in network policies, and it is derived from different layer of defense. This method is clearly semantics in policy configuration and as well as update mechanism to indicate a unique framework security policy. Fig. 7 is presented the variety of policy rules in terms of portions at updating mechanism. This method greatly gives the ease of administrative of the framework and minimizes the probability of the mis-configuration in the security policy.

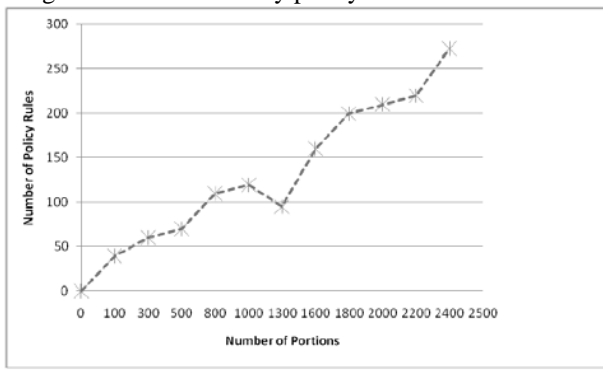

Fig. 7 Number of DPFF Policy Rules in terms of Portions

## 4.6 DPFF Policy Updating

To the view of firewall activity, updating is a complicated procedure to keep the firewall always in development part and never end. As new threats, unauthorized IP domicile and protocol activity are carried out, the new patches and updating procedure should response to the evolving insecure activities [1]. In the proposed framework, the DPFF is under control and development to be update through their policies by administrator undertake and examine all updating procedure which applicable to the firewalls. The status of updating mechanism and minimum network bandwidth consumption is shown in Fig. 8.

## 4.7 Log Analyzer

The out come of the prior experiment are initiated in the firewall log (Table III) and the finest technique to performance of the analyzer is to check and analyze the iptables.

To achieve a considerable field in the log file every record of log is analyses and saved separately. Here it should include that merely functional and valuable data is kept separately at it has shown in the Table IV, for further processing that may sustain and useless information is not concern. These significant data aims to packet structure with their assessment and store them individually for addition testing. The Table IV is shown this significant information.

In the same way as the logged packets, are only the decline packets, all the logged packets which are related with the determination are dropped, and the rest packets are allied with the allow verdict. As we have implemented the KMy firewall to aim for analyze [9], and implementation of the update policy and log analyzer the graphic user interface is simply helping for the compression and analyzer monitoring.

## 5. Conclusions and Future Works

Firewalls are utilized typically to be the main layer of security in the network framework. This chapter is presented the particular segment of the proposed framework that DPFF based on the iptable firewall to be the layers of defense, which is protected front and backend of the framework with a dynamic security and policy update to control the framework's safeguard. A firewall policy commands how the network traffic bypasses and handles by firewall and traffic applications handled. The applicable policy also illustrates the firewall updating and restriction. Establish of firewall policy is to support the traffic application, and establish of firewall rules based on the IP domicile, ports and protocol.

Table 3: Sample of Collected Log

Date: May 25  Time:03:19:01 DENY portmap IN=eth0 SRC=172.168.0.4 DST = 96.17.182.18
PROTO = TCP SPT = 49634 DPT = 80 ACK

Date: May 25  Time:03:19:17      ACCEPT      portmap      IN=eth0      SRC=172.168.0.4
DST=74.123.236.72 PROTO = TCP SPT = 49624 DPT = 80 ACK

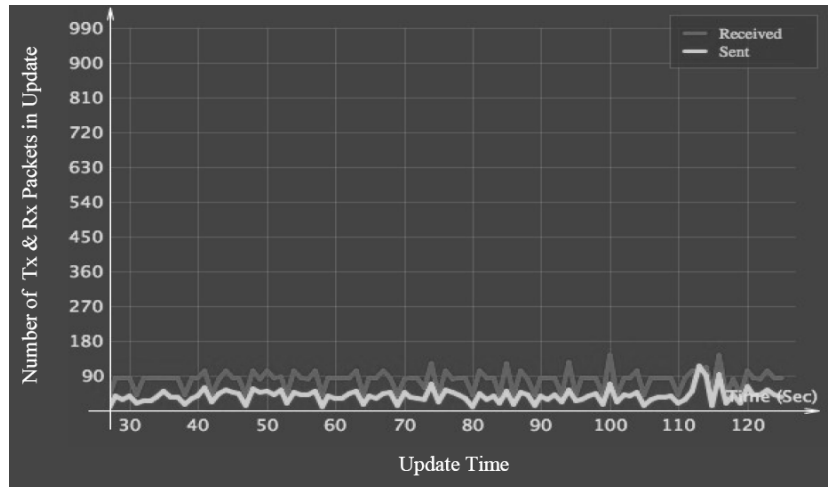

Fig. 8 Analyze of Rules Updating

Table 4: Require Data From Collected Log

|  | IP Domicile | Port | Protocol |
|---|---|---|---|
| Source | 172.168.0.4 | 49624 | TCP |
| Target | 74.125.236.72 | 80 | |

The update mechanism has the following characteristics that, new policy is formed and appended at the initiation of the existing policy. New updated policy is created without almost any similar rules. After the firewall updating and new configuration, the proposed implemented firewall has the distinctiveness that the firewall policies rules are based on the defined and develop rules' manage the firewall to be utilized.

For accuracy in detection and removing possible mis-configuration from the updated policy, it seems rectification algorithms, which determine potential errors, and also investigation in redundancy and shadowing is required.

**Harleen Kaur** gained her Ph.D. in Computer Science from Jamia Millia Islamia University, New Delhi, India on the topic of Applications of Data Mining techniques in Health care Management. She graduated from the University of Delhi, New Delhi. She has previously served as a Lecturer in Computer Science, University of Delhi. Currently, she is an Assistant Professor at the Department of Computer Science, Hamdard University. She has published numerous research articles in refereed international journals and conference proceedings and chapters in an edited book. She is a member of several international bodies. Her main research interests are in the fields of Data analysis with applications to medical databases, Medical decision making, Fuzzy logic, Information Retrieval, Bayesian networks and visualization.

**Omid Mahdi Ebadati Esfahani** is a senior Ph.D. Scholar in Hamdard University, New Delhi, India. He received his MSc degree in Computer Science from Hamdard University, New Delhi, India with top student academic award. He has published numerous international research papers in conferences and journals in computer networks field. He is a member of IEEE, IEEE Computer Society, IACSIT and currently reviewer of different IEEE conferences and peer-reviewed journals. His research interest includes computer networks, network security and Wireless Sensor Networks.

**M. Afshar Alam** is a Professor in Computer Science and former Head, Department of Computer Science, former Dean Faculty of Management and Information Technology, at the Hamdard University, New Delhi, India. In 1997-2000, he founded the Department of Computer Science, Hamdard University. He was also founder of Computer Centre at Hamdard University. He received his Master degree in Computer Application from the Aligarh Muslim University, Aligarh and Ph.D. from Jamia Millia Islamia University, New Delhi. His research interests include Fuzzy logic, Software engineering, Networking and Bioinformatics. He is the author of over 60 publications in International/ National journals, conference and chapter in an edited book. He is a member of expert committee AICTE, DST, UGC and Ministry of Human Resource Development (MHRD), New Delhi, India. His latest books are: Application Software Reengineering, Pearson Education India, 2010, and Advanced Signal Analysis and Its Applications to Mathematical Physics, IK International Publishing House, 2009.